\documentstyle[12pt]{article}

\font\twlgot =eufm10 scaled \magstep1
\font\egtgot =eufm8
\font\sevgot =eufm7

\font\twlmsb =msbm10 scaled \magstep1
\font\egtmsb =msbm8
\font\sevmsb =msbm7

\newfam\gotfam
\def\pgot{\fam\gotfam\twlgot}
\textfont\gotfam\twlgot
\scriptfont\gotfam\egtgot
\scriptscriptfont\gotfam\sevgot
\def\got{\protect\pgot}

\newfam\msbfam
\textfont\msbfam\twlmsb
\scriptfont\msbfam\egtmsb
\scriptscriptfont\msbfam\sevmsb
\def\Bbb{\protect\pBbb}
\def\pBbb{\relax\ifmmode\expandafter\Bb\else\typeout{You cann't use
Bbb in text mode}\fi}
\def\Bb #1{{\fam\msbfam\relax#1}}

\def\thebibliography#1{\bigskip\section*{\large
\bf References\\}\list
  {[\arabic{enumi}]}{\settowidth\labelwidth{#1}\leftmargin\labelwidth
    \advance\leftmargin\labelsep
    \usecounter{enumi}}
    \def\newblock{\hskip .11em plus .33em minus .07em}
    \sloppy\clubpenalty4000\widowpenalty4000
    \sfcode`\.=1000\relax}

\def\op#1{\mathop{{\it\fam0} #1}\limits}

\newcommand{\hm}{{\rm Hom\,}}
\newcommand{\dif}{{\rm Diff\,}}

\newcommand{\bll}{\bullet}
\newcommand{\beq}{\begin{equation}}
\newcommand{\eeq}{\end{equation}}
\newcommand{\ben}{\begin{eqnarray}}
\newcommand{\een}{\end{eqnarray}}
\newcommand{\be}{\begin{eqnarray*}}
\newcommand{\ee}{\end{eqnarray*}}
\newcommand{\bea}{\begin{eqalph}}
\newcommand{\eea}{\end{eqalph}}

\newcommand{\cA}{{\cal A}}
\newcommand{\cJ}{{\cal J}}

\newcommand{\cI}{{\cal I}}

\newcommand{\cZ}{{\cal Z}}

\newcommand{\cK}{{\cal K}}

\newcommand{\dl}{\delta}

\newcommand{\m}{\mu}

\newcommand{\bb}{{\bf 1}}

\newcommand{\ol}{\overline}

\newcommand{\ar}{\op\longrightarrow}

\newcommand{\ot}{\otimes}

\newenvironment{eqalph}{\stepcounter{equation}
\setcounter{equationa}{\value{equation}}
\setcounter{equation}{0}

\begin{eqnarray}}{\end{eqnarray}\setcounter{equation}{\value{equationa}}}

\newcounter{example}
\newcounter{remark}
\newcounter{theorem}
\newcounter{proposition}
\newcounter{lemma}
\newcounter{corollary}
\newcounter{definition}

\def\theremark{\arabic{remark}}

\def\thedefinition{\arabic{definition}}

\newenvironment{theo}{\refstepcounter{definition} \medskip\noindent{\bf
Theorem \thedefinition.}}{\medskip }

\newenvironment{defi}{\refstepcounter{definition} \medskip\noindent{\bf 
Definition \thedefinition.} }{\medskip }

\newcommand{\mar}[1]{}

\begin{document}
\hbox{}

{\parindent=0pt

{\large \bf Jets of modules in 
noncommutative geometry}  
\bigskip 

{\bf G. Sardanashvily}

\medskip

\begin{small}

Department of Theoretical Physics, Moscow State University, 117234
Moscow, Russia

E-mail: sard@grav.phys.msu.su

URL: http://webcenter.ru/$\sim$sardan/
\bigskip

{\bf Abstract.}
Jets of modules over a commutative ring are well known to make up the
representative objects of linear differential operators on these modules.
In noncommutative geometry, jets of modules provide the representative
objects only of a certain class of first order differential operators. 
As a consequence, a generalization of 
the standard Lagrangian formalism on smooth manifolds to noncommutative 
spaces is problematic. 
\end{small}
}

\bigskip
\bigskip

Let $\cK$ be a commutative ring and $\cA$ a commutative $\cK$-ring.
Let $P$ and $Q$ be $\cA$-modules and $\cJ^r(P)$ the module of $r$-order
jets of $P$.
There is a $\cA$-module isomorphism
\mar{cc1}\beq
\hm_\cA(\cJ^r(P),Q)=\dif_r(P,Q), \label{cc1}
\eeq
where $\dif_r(P,Q)$ is the module of $r$-order $Q$-valued linear
differential operators on an $\cA$-module $P$ \cite{kras}. It follows
that the jet module $\cJ^r(P)$ is the representative object of the functor
$Q\to \dif_r(P,Q)$.

In
particular, let $\cA=C^\infty(X)$ be the $\Bbb R$-ring of smooth real
functions on a smooth manifold $X$. If $P$ is a projective
$C^\infty(X)$-module of finite rank, it is isomorphic to the module
$Y(X)$ of
global sections of some vector bundle $Y\to X$. In this case, 
$\cJ^r(P)$ is the module of global sections of the $r$-order jet bundle
$J^rY\to X$ of $Y\to X$. Furthermore, let $Y\to X$ be an arbitrary
smooth fibre bundle. An $r$-order (not necessarily linear) differential
operator on $Y\to X$ is defined as a bundle morphism of $J^rY\to X$ to
some smooth fibre bundle over $X$. As a consequence, Lagrangian
formalism on smooth manifolds is conventionally phrased in the jet terms.

If $\cA$ is not commutative, there are different definitions of 
(linear) differential operators on modules over $\cA$ 
\cite{bor97,dublmp,jara,lunts,epr}. Though the notion of a jet can 
be extended to
a module $P$ over a noncommutative ring $\cA$, only the first order jet
module $\cJ^1(P)$ 
is the representative object of a certain class of first order
differential operators on $P$. 

Let $\cA$ be a
commutative $\cK$-ring. Let $P$ and $Q$ be $\cA$-modules (central bimodules).
The $\cK$-module $\hm_\cK (P,Q)$
of $\cK$-linear homomorphisms $\Phi:P\to Q$ is endowed with the
two different $\cA$-module structures
\mar{5.29}\beq
(a\Phi)(p):= a\Phi(p),  \qquad  (\Phi\bll a)(p) := \Phi (a p),\qquad a\in
\cA, \quad p\in P. \label{5.29}
\eeq
We will refer to the second one as the $\cA^\bll$-module structure.
Let us put
\mar{spr172}\beq
\dl_a\Phi:= a\Phi -\Phi\bll a, \qquad a\in\cA. \label{spr172}
\eeq
The following definitions are equivalent.

\begin{defi} \label{ws131} \mar{ws131}
An element $\Delta\in\hm_\cK(P,Q)$ is an $r$-order $Q$-valued  
differential operator on $P$ if
$\dl_{a_0}\circ\cdots\circ\dl_{a_r}\Delta=0$
for any tuple of $r+1$ elements $a_0,\ldots,a_r$ of $\cA$.
\end{defi}

\begin{defi} \label{ws155} \mar{ws155}
An element $\Delta\in \hm_\cK (P,Q)$ is a zero order differential
operator if $\dl_a\Delta=0$ for all $a\in\cA$, and $\Delta$ is 
a differential operator of order $r>0$ if $\dl_a\Delta$ for all $a\in\cA$
is an $(r-1)$-order differential operator.
\end{defi}

Given an $\cA$-module $P$, let us consider the tensor product
$\cA\otimes_\cK P$ of $\cK$-modules $\cA$ and $P$.
We put
\mar{spr173}\beq
\dl^b(a\otimes p):= (ba)\otimes p - a\otimes (b p), \qquad p\in
P, \qquad a,b\in\cA.  \label{spr173}
\eeq
Let us denote by $\m^{k+1}$ the $\cA$-submodule of
$\cA\ot_\cK P$ generated by elements of the type
\be
\dl^{b_0}\circ \cdots \circ\dl^{b_k}(\bb\otimes p).
\ee

\begin{defi} \label{cc2}
The $k$-order jet module $\cJ^k(P)$ of a
module $P$ is defined as the quotient of the $\cK$-module $\cA\otimes_\cK
P$ by $\m^{k+1}$.
We denote its elements $a\ot_kp$.
\end{defi}

The $\cK$-module $\cJ^k(P)$ is endowed with the $\cA$- and $\cA^\bll$-module
structures
\mar{+a21}\beq
b(a\ot_k p):= ba\ot_k p, \qquad
b\bll(a\otimes_k p):= a\otimes_k (bp). \label{+a21}
\eeq
There exists the module morphism
\mar{5.44}\beq
J^k: P\ni p\to \bb\otimes_k p\in \cJ^k(P)
\label{5.44}
\eeq
of the $\cA$-module $P$ to the $\cA^\bll$-module $\cJ^k(P)$ such that
$\cJ^k(P)$, seen as an $\cA$-module, is generated by
elements $J^kp$, $p\in P$.

\begin{theo} \label{t6} \mar{t6}
Any $Q$-valued differential operator $\Delta$ of order $k$ on an
$\cA$-module $P$ 
factorizes uniquely 
\be
\Delta: P\ar^{J^k} \cJ^k(P)\ar Q
\ee
through the morphism $J^k$ (\ref{5.44}) and some $\cA$-module
homomorphism ${\got f}^\Delta: \cJ^k(P)\to Q$ \cite{kras}.
\end{theo}

Let us denote $J: P\ni p\mapsto \bb\ot p\in \cA\ot P$.
The proof is based on the fact that
\mar{cc3}\beq
\dl_{b_0}\circ \cdots \circ\dl_{b_k}({\got f}\circ J)(p)
={\got f}(\dl^{b_0}\circ \cdots \circ\dl^{b_k}
(\bb\ot p)) \label{cc3}
\eeq
for any ${\got f}\in\hm_\cA (\cA\ot P,Q)$.
The correspondence $\Delta\mapsto {\got f}^\Delta$ yields the above
mentioned isomorphism (\ref{cc1}).

Let now a $\cK$-ring $\cA$ need not be commutative. 
Let $P$ and $Q$ be two-sided $\cA$-modules. Two-sided $\cA$-modules
throughout are assumed to be central bimodules over the center
$\cZ_\cA$ of $\cA$.  The $\cK$-module $\hm_\cK(P,Q)$
can be provided with the left $\cA$- and right $\cA^\bll$-module structures
(\ref{5.29}) and the similar right and left structures
\mar{ws105}\beq
(\Phi a)(p):=\Phi(p)a, \qquad (a\bll\Phi)(p):=\Phi(pa), \qquad a\in\cA, \qquad
p\in\ P. \label{ws105}
\eeq
For the sake of convenience, we will refer to $\cA-\cA^\bll$ structures
(\ref{5.29}) and (\ref{ws105}) as the left and right $\cA-\cA^\bll$
structures, respectively. 
Let us put 
\mar{ws133}\beq
\ol\dl_a\Phi:=\Phi a-a\bll\Phi, \qquad a\in\cA,
\qquad \Phi\in \hm_\cK(P,Q). \label{ws133}
\eeq
It is readily observed that  
$\dl_a\circ\ol\dl_b=\ol\dl_b\circ\dl_a$ for all $a,b\in\cA$.

We follow the notion of a differential operator in \cite{lunts}.
Let $P$ and $Q$ be regarded as
left $\cA$-modules. 
Let us consider the $\cK$-module $\hm_\cK (P,Q)$ provided with the left
$\cA-\cA^\bll$ module structure (\ref{5.29}).
We denote $\cZ_0$ its center, i.e., $\dl_a\Phi=0$ for all $\Phi\in\cZ_0$
and $a\in\cA$. Let $\cI_0=\ol \cZ_0$ be the $\cA-\cA^\bll$ submodule 
of $\hm_\cK (P,Q)$ 
generated by $\cZ_0$. Let us consider: (i) the quotient $\hm_\cK (P,Q)/\cI_0$,
(ii) its center $\cZ_1$, (iii) the $\cA-\cA^\bll$ submodule $\ol \cZ_1$
of $\hm_\cK (P,Q)/\cI_0$ generated by $\cZ_1$, and (iv) the
$\cA-\cA^\bll$ submodule $\cI_1$  
of $\hm_\cK (P,Q)$ given by the relation 
$\cI_1/\cI_0=\ol \cZ_1$.
Then we
define the $\cA-\cA^\bll$ submodules $\cI_r$, $r=2,\ldots$, of
$\hm_\cK (P,Q)$ by induction
as follows:
\mar{ws134}\beq
\cI_r/\cI_{r-1}=\ol \cZ_r, \label{ws134}
\eeq
where $\ol \cZ_r$ is the $\cA-\cA^\bll$ module 
generated by the center $\cZ_r$ of the quotient $\hm_\cK (P,Q)/\cI_{r-1}$.
 
\begin{defi} \label{ws135} \mar{ws135}
Elements of the submodule $\cI_r$ of $\hm_\cK (P,Q)$ are said to be
the left $r$-order $Q$-valued  
differential operators on a two-sided $\cA$-module $P$.
\end{defi}

An equivalent definition is the following \cite{epr}

\begin{defi} \label{ws137} \mar{ws137}
An element of $\Delta\in\hm_\cK (P,Q)$ is a left
$r$-order $Q$-valued  
differential operator on an $\cA$-module $P$ if
it is a finite sum 
\mar{ws138}\beq
\Delta(p)=b_i\Phi^i(p) +\Delta_{r-1}(p), \qquad b_i\in\cA, \label{ws138}
\eeq
where $\Delta_{r-1}$ and $\dl_a\Phi^i$ for all $a\in\cA$ are left 
$(r-1)$-order 
differential operators if $r>0$, and they vanish if $r=0$.
\end{defi}

The set of these operators
is provided with the left and
right $\cA-\cA^\bll$ module structures.

If $\cA$ is a commutative ring,
Definitions \ref{ws135} and \ref{ws137} come to Definition \ref{ws155}. 
  
If $P=Q=\cA$, derivations of $\cA$ and their compositions are differential
operators on $\cA$ in accordance with Definition \ref{ws137} \cite{epr}.

By analogy with Definitions \ref{ws135} and \ref{ws137}, one
can define right differential operators on two-sided $\cA$-modules as follows.

\begin{defi} \label{ws151} \mar{ws151}
Let $P$ and $Q$ be seen as right $\cA$-modules over a 
noncommutative $\cK$-ring $\cA$.
An element $\Delta\in\hm_\cK(P,Q)$ is said to be a right zero order $Q$-valued 
differential operator on $P$ if it is a finite sum
$\Delta=\Phi^i b_i$, $b_i\in\cA$, 
where $\ol\dl_a\Phi^i=0$ for all $a\in\cA$. An element $\Delta\in\hm_\cK(P,Q)$
is called a right
differential operator of order $r>0$ on $P$ if
\mar{ws150}\beq
\Delta(p)=\Phi^i(p)b_i +\Delta_{r-1}(p), \qquad b_i\in\cA, \label{ws150}
\eeq
where $\Delta_{r-1}$ and $\ol\dl_a\Phi^i$ for all $a\in\cA$ are right 
$(r-1)$-order differential operators.
\end{defi}

Definition \ref{ws135} and Definition
\ref{ws151} of differential operators on two-sided $\cA$-modules are
not equivalent, but one can combine them as follows. 

\begin{defi} \label{ws152} \mar{ws152}
Let $P$ and $Q$ be two-sided modules over a noncommutative $\cK$-ring $\cA$.
An element $\Delta\in\hm_\cK(P,Q)$ is a two-sided zero order $Q$-valued 
differential operator on $P$ if 
it is either a left or right zero order differential operator.
An element $\Delta\in\hm_\cK(P,Q)$ is said to be a two-sided differential
operator of order $r>0$ on $P$ if 
it is brought both into the form $\Delta=b_i\Phi^i +\Delta_{r-1}$, $b_i\in\cA$,
and 
$\Delta=\ol\Phi^i\ol b_i +\ol\Delta_{r-1}$, $\ol b_i\in\cA$,
where $\Delta_{r-1}$, $\ol\Delta_{r-1}$ and
$\dl_a\Phi^i$, $\ol\dl_a\ol\Phi^i$ for all $a\in\cA$  
 are two-sided $(r-1)$-order differential operators.
\end{defi}

One can think of this definition as a generalization of
the definition of first order differential operators in noncommutative
geometry in \cite{dublmp}.

It is readily observed that two-sided differential operators in
Definition \ref{ws152} need not be left or right differential
operators, and {\it vice versa}.   
At the same time, $\cA$-valued derivations of a $\cK$-ring $\cA$ and their
compositions obey Definition \ref{ws152}.

Turn now to the notion of jets in noncommutative geometry. One can
follow Definition \ref{cc2} in order to define the left jets of a
two-sided module $P$ over a noncommutative ring $\cA$. However, the
relation (\ref{cc3}) does not hold if $k>0$. Therefore, no module of
left jets is the representative object of left differential operators.
The notion of a right jet of $P$ meets the similar problem.

Given an $\cA$-module $P$, let us consider the tensor product
$\cA\otimes_\cK P\ot_\cK\cA$ of $\cK$-modules $\cA$ and $P$.
We put
\be
&& \dl^b(a\otimes p\ot c):= (ba)\otimes p\ot c - a\otimes (bp)\ot c,\\
&& \ol\dl^b(a\otimes p\ot c):= a\otimes p\ot (cb) - a\otimes (pb)\ot c,
\qquad p\in P, \qquad a,b,c\in\cA.
\ee
Let us denote by $\m^1$ the two-sided $\cA$-submodule of
$\cA\ot_\cK P\ot_\cK\cA$ generated by elements of the type
$\ol\dl^c\circ\dl^b(\bb\otimes p\ot\bb)$.
One can define the first order two-sided jet module $\cJ^1(P)=\cA\otimes_\cK
P\ot_\cK\cA/\m^1$ of 
$P$. Let us denote 
\be
J: P\ni p\mapsto \bb\ot p\ot\bb\in \cA\ot P\ot\cA.
\ee
Then the equality 
\be
\dl_{b_0}\circ \cdots \circ\dl_{b_k}({\got f}\circ J)(p)
={\got f}(\dl^{b_0}\circ \cdots \circ\dl^{b_k}
(\bb\ot p\ot\bb)) 
\ee
holds for any ${\got f}\in\hm_{\cA-\cA} (\cA\ot P,Q)$.
It follows that $\cJ^1(P)$ is the representative object of the functor
$Q\to \ol\dif_1(P,Q)$ where the $\cK$-module $\ol\dif_1(P,Q)$ consists
of two-sided first order differential operators $\Delta$ by
Definition \ref{ws152} which obey the condition
$\ol\dl^c\circ\dl^b\Delta=0$ for all $c,b\in\cA$. They are the
first order differential operators defined in \cite{dublmp}.

\end{document}